\documentstyle[epsfig,sprocl]{article}

\newcommand{\expl}{\langle \!\langle}
\newcommand{\expr}{\rangle \!\rangle}

\begin{document}

\title{LANGEVIN INTERPRETATION
\\
OF KADANOFF-BAYM EQUATIONS }

\author{C. GREINER \footnote{Talk presented at the
Workshop on `Kadanoff-Baym Equations: Progress and Perspectives
for Many-Body Physics',
20.-24. September 1999, Rostock, Germany}, S. LEUPOLD}

\address{Institut f\"ur Theoretische Physik, Universit\"at Giessen,\\
D-35392 Giessen, Germany }

\maketitle\abstracts{
We show that the nonperturbative quantum
transport equations, the `Kadanoff-Baym equations',
can be be understood as the ensemble average over stochastic equations of
Langevin type.
For this we couple a free scalar boson quantum field to
an environmental heat bath with some given temperature $T$.
The inherent presence
of noise and dissipation related by the fluctuation-dissipation theorem
guarantees that the modes or particles become thermally populated on average
in the long-time limit.
This interpretation leads to a more intuitive physical picture of the
process of thermalization and of the interpretation of the Kadanoff-Baym
equations.
}

\section{Motivation} \label{sec:intro}

Non-equilibrium many body theory had been traditionally
a major topic of research for describing various (quantum) transport
phenomena in plasma physics, in condensed matter physics and nuclear physics.
Over the last years
a lot of interest for non-equilibrium quantum field theory has now emerged
also in particle physics.
A very powerful diagrammatic tool is given
by the `Schwinger-Keldysh' or `closed time path' (CTP)
technique by means of non-equilibrium Green's functions for describing a quantum system
also beyond thermal equilibrium.
The resulting causal and nonperturbative equations of motion
(by various approximations),
the so called {\em Kadanoff-Baym (KB) equations}, have to be considered
as an ensemble average over the initial density matrix
characterizing the preparation of the initial state of the system.
If the system behaves
dissipatively, as a consequence of the
famous fluctuation-dissipation theorem, there must exist fluctuations.
The Kadanoff-Baym equations have then to be understood
as an ensemble average over all the possible fluctuations.
This inherent stochastic aspect of the KB equations is what
we want to point out and thus provide, as we believe, some new physical insight
into its merely complex structure \cite{GL98}.
In what follows below we want to point out its intimate connection to
Langevin like processes.

As an elementary reminder of a Langevin process let us first briefly review
the description of classical Brownian motion.
Consider
a heavy `Brownian' particle with mass $M$ placed in a thermal environment
obeying an effective Langevin equation, i.e.
\begin{equation}
M \ddot{x}  + 
2\int\limits_{-\infty}^{t} dt' \, \Gamma (t-t') \, \dot{x}(t') 
=  \xi (t)    \, .
\label{Lang2}
\end{equation}
Here $\xi (t)$ has to be interpreted as a `noisy'
source driving the fluctuations of the Brownian particle.
For many applications $\xi (t)$ is completely
specified by a Gaussian distribution
with zero mean and the correlation kernel $I$ :
\begin{equation}
I(t-t')  :=  \expl \xi(t) \xi(t')\expr
\, \equiv \,  2  T \Gamma (t-t')   \, .
\label{BMnoise1}
\end{equation}
$\expl \ldots \expr$ denotes the average over all possible realizations of the
stochastic variable $\xi (t)$.
In fact, the simple relation between the
dissipation kernel $\Gamma $ and the strength $I$ of the
random force $\xi (t)$ just stated is a
manifestitaion of the fluctuation-dissipation theorem and
(in the long time limit) is in accordance with the
equipartition condition
$\expl p^2 \expr/(2M) \stackrel{!}{=} T/2$.

For a further physical motivation let us return
to quantum field theory and already point out some similarities.
One of the major present topics in quantum field theory at finite temperature
or near thermal equilibrium concerns the evolution and behavior of the long
wavelength modes. These modes often lie entirely in the non-perturbative regime.
Therefore solutions of the classical field equations in Minkowski space have
been widely used in recent years to describe long-distance properties
of quantum fields that require a non-perturbative analysis.
A justification of the classical treatment of the long-distance dynamics
of bosonic quantum
fields at high temperature is based on the observation that the average 
thermal amplitude of low-momentum modes is large.
For the low-momentum modes $|\vec{p }|\ll T$
(and for a weakly coupled quantum field theory) their (Bose) occupation number
$n_B$
approaches the classical equipartition limit.
The classical field equations
should provide a good approximation for
the dynamics of such highly occupied modes.
However, in a correct
semi-classical treatment of the soft modes the hard, i.e. thermal modes
cannot simply be neglected, but it
should incorporate their influence in a consistent way.
In a recent paper \cite{CG97} it was shown how to construct
an effective semi-classical action for describing not only the classical
behavior of the long wavelength modes below some appropriate
cutoff $k_c$,
but taking into account also perturbatively the interaction among the soft
and hard modes. By integrating out the `influence' of the hard modes
on the two-loop level (for standard $\phi^4$-theory) the emerging semi-classical
equations of motion
for the soft fields
can be derived from an effective action
and become stochastic equations of motion
of generalized Langevin type \cite{CG97},
which resemble in their structure
the analogous expression to (\ref{Lang2}).
The hard modes act as an
environmental heat bath. They also guarantee that the soft modes become, on average,
thermally populated with the same temperature as the heat bath.
For the semi-classical regime, where $|\vec{ p}|\ll T$, one finds
for the ensemble average of the squared amplitude
\begin{eqnarray}
\frac{1}{V} \expl |\phi (\vec{ p})|^2 \expr
&  \approx &
\frac{1}{E_p^2} \, T
\approx 
\frac{1}{E_p} \, n_B(E_p)
\, .
\label{equi2}
\end{eqnarray}
Such kind of Langevin description for
the non-perturbative evolution of (super-)soft modes
(on a scale of $|\vec{p}| \sim g^2 T \ll T$)
in non Abelian gauge theories
has recently been put forward \cite{B99}.
The understanding of the behavior of the soft modes is crucial
e.g.~for the issue of 
anomalous baryon number violation
due to the diffusion of topological
Chern-Simons charge
in hot electroweak theory (see eg \cite{B99} and references listed therein).

In analogy to the Langevin description stated above
we want to sketch in the following (pedagogical) study
the effect of the heat bath on
the evolution of the system degrees of freedom by means of the
`closed time path Green's function' (CTPGF) technique.
For this we discuss a free scalar field theory
interacting with a heat bath.

\section{Stochastic Interpretation of KB equations } \label{sec:CTPGF}

\begin{figure}
\centerline{\epsfxsize=4cm \epsfbox{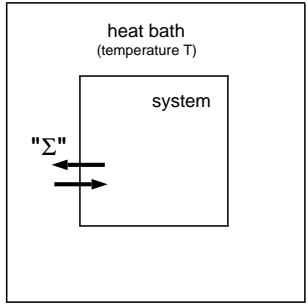}}
\caption{Schematic picture for the interaction of the system with the heat bath.}
\label{figa2}
\end{figure}

We start with the CTP action for a scalar
bosonic field $\phi$ coupled to an environmental heat bath of temperature
$T$:
\begin{eqnarray} 
S &=& 
\int \!\!d^4\!x
{1\over 2} \left[ 
\phi^+ \, (-\Box-m^2) \,\phi^+ - \phi^- \, (-\Box-m^2) \,\phi^- 
\right]
\label{eq:ctpac}
\\ &&
(-) \int \!\!d^4\!x_1
\!d^4\!x_2
{1\over 2} \left[ 
 \phi^+ \,\Sigma^{++} \,\phi^+ + \phi^+ \,\Sigma^{+-} \,\phi^-
+ \phi^- \,\Sigma^{-+} \,\phi^+ + \phi^- \,\Sigma^{--} \,\phi^-
\right]
\,.
\nonumber
\end{eqnarray}
The system starts to evolve from some initial density matrix.
The interaction among the system and the heat bath is stated by
an interaction kernel involving a self energy operator
$\Sigma $ resulting effectively from integrating out the heat bath degrees
of freedom.
Schematically this is sketched in fig.~\ref{figa2}.
In (\ref{eq:ctpac}) the self energy contribution from the heat bath
is parametrized in the Keldysh notation by the four self energy parts, which
can be expressed by means of the standard contributions $\Sigma ^<$
and $\Sigma ^>$.
Clearly, this self energy operator is the only quantity which might drive
the system towards equilibrium.
If the heat bath is at equilibrium, then the
Kubo-Martin-Schwinger relation holds:
\begin{equation}
  \bar \Sigma^>(k) = e^{k_0/T}\, \bar \Sigma^<(k) \,.  \label{eq:heat}
\end{equation}
(Here and in the following $\bar A(k) := \bar A (k_0,\vec{k})$ denotes
the 4-dim. Fourier transform
of a (stationary and translational invariant)
quantity $A(x_1, x_2) \rightarrow A(x=x_1-x_2)$ with respect to
the 4-dim. relative coordinate $x$.)

From (\ref{eq:ctpac}) it is now straightforward
to obtain the equations of motion for 
the characteristic two-point functions \cite{GL98}.
For the retarded propagator one has
\begin{equation}
  \label{eq:eomretprop}
 (-\Box -m^2-\Sigma^{\rm ret}) D^{\rm ret} = \delta \, ,
\end{equation}
where $\Sigma^{\rm ret} :=\Theta(t_1-t_2)\,\left[ \Sigma^> - \Sigma^< \right] $.
Additional dynamical information comes from the equation of motion of
the propagator $D^<$
\begin{equation}
  \label{eq:kadbaym}
(-\Box -m^2)  D^< - \Sigma^{\rm ret} D^< - \Sigma^< D^{\rm av} = 0 \,.
\end{equation}
This is just the famous KB equation.
(\ref{eq:eomretprop}) and (\ref{eq:kadbaym}) determine the complete and causal
(non-equilibrium) evolution for the two-point functions.

To get more physical insight into the (effective) action (\ref{eq:ctpac})
and in the equations of motion we
introduce the following {\em real} valued quantities:
\begin{eqnarray}
  s(x_1,x_2) & := & {1\over 2} {\rm sgn}(t_1-t_2) 
\left(\Sigma^>(x_1,x_2) - \Sigma^<(x_1,x_2) \right) 
= s(x_2,x_1) \, ,
\label{eq:sdef} \\
a(x_1,x_2) & := & {1\over 2} \left(\Sigma^>(x_1,x_2) - \Sigma^<(x_1,x_2) \right) 
= - a(x_2,x_1) \, ,
\label{eq:adef}
\\
I(x_1,x_2) & := & -{1\over 2i} \left(\Sigma^>(x_1,x_2) + \Sigma^<(x_1,x_2) 
\right)  
 =   I(x_2,x_1)
\,.
\label{eq:Idef}
\end{eqnarray}
Our notion for $s$ and $a$ serves as a reminder for the respective symmetry 
properties. It basically represents the standard decomposition of the real
and imaginary part of the Fourier transform of the retarded self energy
operator $  \bar \Sigma^{\rm ret}$.
$s$ yields a (dynamical) {\em mass shift} for the $\phi$
modes caused by the
interaction with the modes of the heat bath, while $a$ is responsible for the
damping, i.e. {\em dissipation}
of the $\phi$ fields. The important thing to point out will be that
$I$ characterizes the {\em fluctuations}.

We first note that the CTP action (\ref{eq:ctpac})
can be written as
\begin{eqnarray}
  S & = &
\int \!\!d^4\!x
{1\over 2} \left[ 
\phi^+ \, (-\Box-m^2) \,\phi^+ - \phi^- \, (-\Box-m^2) \,\phi^- 
\right]
\label{eq:IFac}   
\\
&&
\int \!\!d^4\!x_1
\!d^4\!x_2
{1\over 2} \left[ 
- (\phi^+ - \phi^-)\, (s+a) \, (\phi^+ + \phi^-)
+ i \, (\phi^+ - \phi^-) \, I \, (\phi^+ - \phi^-) 
\right]  \,.
\nonumber
\end{eqnarray}
This expression is
identical to the so called {\em influence functional}
given by Feynman and Vernon.
To the exponential factor $e^{iS}$ in the path integral
the $(s+a)$ term contributes a phase while the $I$ term
causes an exponential damping and thus signals nonunitary evolution.

The two relevant equations of motion are stated as
\begin{eqnarray}
  \label{eq:eomretsa}
(-\Box -m^2-s-a) D^{\rm ret}  & = & \delta \,, \\
  \label{eq:eomklsaI}
(-\Box -m^2-s-a)  D^< + (a+iI) D^{\rm av} & = & 0 \,.
\end{eqnarray}
We see that the last equation
is the only one where $I$ occurs.

For the interpretation of $s$, $a$ and $I$
consider the long-time behavior of these equations.
In this case we can assume that the system becomes translational
invariant in time and space and the boundary terms are no longer important. 
For the spectral function one immediately finds
\begin{equation}
{\cal A}(k)  :=
{i\over 2} [\bar D^{\rm ret}(k) - \bar D^{\rm av}(k)] 
 =  {i\, \bar a(k) \over [k^2-m^2-\bar s(k)]^2 + \vert \bar a(k) \vert ^2 } \,.
\label{eq:defspec}
\end{equation}
It becomes obvious that
$\bar s\equiv \Re \bar \Sigma ^{\rm ret}$
contributes an 
(energy dependent) {\em mass shift} while
$\bar a \equiv \Im \bar \Sigma ^{\rm ret}$
causes the {\em damping} of propagating
modes. $\bar a$ is related to the commonly used
damping rate $\bar{ \Gamma}$ via
\begin{equation}
  \label{defdamprate}
\bar{ \Gamma} (k)  =  i \frac{\bar{a}(k)}{k_0}   \, .
\end{equation}
For $D^<$ one finds in the long-time limit the relation
\begin{eqnarray}
\bar D^<(k) &= & \bar D^{\rm ret}(k)
\bar \Sigma^{<} \bar D^{\rm av}(k)  \nonumber  \\
&= & \bar D^{\rm ret}(k)
[-\bar a(k)-i\bar I(k)] \bar D^{\rm av}(k)
\equiv	 -2i \,n(k) {\cal A}(k)  \, ,
  \label{eq:dkltherm}
\end{eqnarray}
where, by employing KMS condition (\ref{eq:heat}),
\begin{equation}
  \label{eq:finpnd}
n(k)  = 
{\bar \Sigma^<(k) \over \bar\Sigma^>(k) -\bar\Sigma^<(k) } 
= {1 \over e^{k_0/T} -1 }
 \equiv  n_B(k_0)  \, ,
\end{equation}
which indeed shows that the phase space occupation number in
the long-time limit becomes a Bose distribution with the temperature
of the heat bath.

It is now very illuminating to explicitly write down the relation between 
$\bar a(k)$ and $\bar I(k)$ using the
definitions (\ref{eq:adef}) and (\ref{eq:Idef}):
\begin{eqnarray}
  \label{eq:fludis}
\bar I(k) = 
{\bar\Sigma^>(k) +\bar\Sigma^<(k) \over \bar\Sigma^>(k) -\bar\Sigma^<(k)} \, 
i \, \bar a(k) = {\rm coth}\!  \left({k_0 \over 2T}\right) i \, \bar a(k) \,. 
\end{eqnarray}
In the high temperature (classical) limit $(k_0\ll T)$ one gets
\begin{equation}
  \label{eq:hight}
  \bar I(k) = {T \over k_0} 2i\, \bar a(k) \,,
\end{equation}
or, employing (\ref{defdamprate}),
\begin{equation}
  \label{eq:hight1}
  \bar I(k) = 2T\, \bar{\Gamma }(k) \,.
\end{equation}
Recalling the discussion of Brownian motion in the introduction
this compares favorably well with (\ref{BMnoise1}).
The physical meaning of $I$ as a `noise' correlator will become obvious.
The relation (\ref{eq:fludis}) thus represents the
{\em generalized fluctuation-dissipation relation}
from a microscopic point of view
by the various definitions of $\bar{I},\,  \bar{a}$ and $\bar{\Gamma }$
through the parts $\bar{\Sigma }^<$ and $\bar{\Sigma }^>$ of the self energy.

To see now more closely the connection
to stochastic equations we decompose the influence action $S$
as given in (\ref{eq:IFac})
in its real and imaginary part and write for the corresponding
generating functional
\cite{GL98}
\begin{eqnarray}
\lefteqn{Z[j^+,j^-]} \nonumber \\
& := & \int\!\! {\cal D}[\phi^+,\phi^-] \,
\rho[\phi^+,\phi^-]\,  
e^{iS[\phi^+,\phi^-]+i\,j^+\phi^+ + i\,j^-\phi^-}
\nonumber \\
& = & \int\!\! {\cal D}[\phi^+,\phi^-] \, 
\rho[\phi^+,\phi^-]\,
e^{i{\Re }S[\phi^+,\phi^-] +i\,j^+\phi^+ + i\,j^-\phi^-
   -{1\over 2} (\phi^+ - \phi^-) \, I \, (\phi^+ - \phi^-) } 
\nonumber \\
& = & {1\over \tilde N} \int\!\! {\cal D}[\xi ] \,
       e^{-{1\over 2}\xi \,I^{-1}\xi } \,
      \int\!\! {\cal D}[\phi^+,\phi^-] \,
\rho[\phi^+,\phi^-]\, 
e^{i{\Re }S[\phi^+,\phi^-] +i\,j^+\phi^+ + i\,j^-\phi^-
   +i\,\xi\,(\phi^+ - \phi^-) }
\nonumber \\
& = & {1\over \tilde N} \int\!\! {\cal D}[\xi ] \,
       e^{-{1\over 2}\xi \,I^{-1}\xi } \,
       Z'[j^+ +\xi ,j^- -\xi ]  \,
\, \equiv \,   \expl Z'[j^+ +\xi,j^- -\xi]\, \expr
  \label{eq:ZZpr}
\end{eqnarray}
with 
$ \tilde N := \int\!\! {\cal D}\xi \,
e^{-{1\over 2}\xi \,I^{-1}\xi } \,$.
The action entering the definition of $Z'$ is no longer $S$, but only the real
part of the influence action (\ref{eq:IFac}).
The generating functional $Z[j^+,j^-]$ can thus be
interpreted as a new stochastic
generating functional $Z'[j^+ +\xi,j^- -\xi]$ averaged over a random Gaussian
(noise) field $\xi$ with the width function $I$,
i.e. 
\begin{equation}
  \label{eq:defavdbl}
\expl {\cal O} \expr := 
{1\over \tilde N}
\int\!\! {\cal D}\xi \, {\cal O}\, e^{-{1\over 2}\xi \,I^{-1}\xi } \,. 
\end{equation}
From the last definition we find that the (ensemble) average
over the noise field vanishes, i.e.
$ \expl \xi \expr = 0 $  ,
while the noise correlator is given by 
\begin{equation}
  \label{eq:noisecor}
\expl \xi \xi \expr = I \, .
\end{equation}

From the stochastic functional $Z'$
a {\em Langevin} equation for a classical $\phi$ field can now readily
be derived.
Noting that the fields
$\langle\phi^+ \rangle_\xi$ on the upper branch and
$\langle\phi^- \rangle_\xi$ on the lower branch are equal
(and denoted as $\phi_\xi$ in the following),
its equation of motion derived from $Z'$ takes the form
\begin{equation}
  \label{eq:eomcl2}
(-\Box-m^2-s) \, \phi_\xi- a \, \phi_\xi= -\xi \,. 
\end{equation}
This, indeed, represents a standard Langevin equation.
The spatial Fourier transform of the Langevin equation (\ref{eq:eomcl2})
then takes the form
\begin{equation}
\ddot{\phi}_\xi (\vec{k},t)
+(m^2+\vec{k}^2-2\Gamma (\vec{k},\Delta t=0)) \phi_\xi
+ 2\int\limits_{-\infty}^{t} dt' \Gamma (\vec{k},t-t')
 \dot{\phi}_\xi (\vec{k},t')
 =  \xi (\vec{k},t)  .
  \label{eq:eomcl3}
\end{equation}
The analogy between this Langevin equation (\ref{eq:eomcl3}) and the
one for a single classical oscillator is obvious.
The important difference, however, is the fact that the
corresponding relations (\ref{eq:fludis}) and (\ref{BMnoise1})
between the respective noise kernel $I$ and friction kernel $\Gamma $
only agree in the high temperature limit.

One can further ask to what extend the classical equations of motion
(\ref{eq:eomcl2}) together with (\ref{eq:noisecor})
are an approximation for the full quantum problem given by the
equation of motion (\ref{eq:eomklsaI}) for $D^<$.
Inverting (\ref{eq:eomcl2}) one finds for the correlation function
in the long-time limit
\begin{eqnarray}
  \label{eq:clapp}
-i\expl \, \langle \phi^+ \rangle_\xi \, \langle \phi^- \rangle_\xi \expr 
= -i \, D^{\rm ret} \expl \xi \xi \expr D^{\rm av} 
= -i \, D^{\rm ret} I\, D^{\rm av} \,.
\end{eqnarray}
Note that (\ref{eq:clapp}) is indeed the relation (\ref{equi2})
advocated in the introduction to hold in the
(semi-)classical regime.
This one has to compare with the full quantum correlation function
$D^<$ of (\ref{eq:dkltherm}).
One thus has that $(-\bar a-i\bar I)$ is approximated by $-i\bar I$.
Of course this is justified, if 
$\vert \bar a \vert \ll \bar I$
holds. Using the microscopic quantum version (\ref{eq:fludis}) of the 
fluctuation-dissipation theorem this is equivalent to
${\rm coth}\left( {k_0 \over 2T} \right) \gg 1 $.
Thus in the high temperature limit or -- turning the argument
around -- for low frequency modes, i.e.
for $k_0 \ll T $,
the classical solution yields a good approximation to the full quantum case.
To be more precise: In simulations one has to solve the classical Langevin 
equation (\ref{eq:eomcl2}) and calculate $n$-point functions by averaging over the
random sources.

One can also write down the equations of motion for the
quantum two-point functions with external noise \cite{GL98} by introducing
'noisy' two-point propagators.
Averaging the equation of motion over the noise fields according to
(\ref{eq:defavdbl}) one indeed rederives the KB equation.
This demonstrates that the KB equation can be interpreted
as an ensemble average over fluctuating fields which are subject to noise, the 
latter being correlated by the sum of self energies $\Sigma^<$ and $\Sigma^>$,
i.e.~from a transport theoretical point of view the sum of production and 
annihilation rate.
We want to note once more that the `noisy'
or fluctuating part denoted by $I$ inherent to the structure of the
KB equation (\ref{eq:kadbaym}) guarantees that the modes
or particles become correctly (thermally) populated, as can be
realized by inspecting (\ref{eq:dkltherm}).

We close our discussion by noting that one can also pursue to
derive a standard kinetic transport equation for the (semi-classical) phase-space distribution
$f(\vec{x}, \vec{k},t)$ including fluctuations \cite{GL98}.
The derived kinetic transport process
has the structure of the phenomenologically inspired
Boltzmann-Langevin equation.
Our approach carried out in \cite{GL98} has to be considered
as a clear derivation from first principles. Indeed it shows (nearly)
a one to one correspondence to the phenomenologically introduced scheme.
However, also some severe interpretational difficulties in the interpretation
of the fluctuating phase-space density remain. We refer the interested reader
to our discussion in \cite{GL98}.

\section{Some further conclusions} \label{sec:summary}

In our discussions we have elucidated
on the stochastic aspects inherent to the (non-) equilibrium quantum transport
equations.
We have isolated a
term denoted by {\em I} which solely characterizes the
(thermal and quantum) fluctuations inherent to the underlying transport process.
By introducing a stochastic
generating functional the emerging stochastic equations of motion
can then be seen as generalized (quantum) Langevin processes.
What is changed, if we replace 
our toy model of a free system coupled to an external heat bath
by a self-coupled and thus nonlinear closed system?
In an interacting field theory of a closed system the KB
equations formally have exactly the same structure as in our toy model. The
important difference, however, is that the self energy operator is now
described fully (within the appropriate approximative scheme) by the
system variables, i.e.~it is expressed as a convolution of various two-point
functions. Hence, an underlying simple stochastic process,
as in our case an external
stochastic Gaussian process, cannot really be extracted. However, we
emphasize that the emerging structure of the KB
equations is identical. The decomposition of the
self energy operator into its three physical parts (mass shift $s$, damping $a$, 
and fluctuation kernel $I$) can immediately be
taken over. Hence these three parts keep their clear physical meaning
also for a nonlinear closed system.

One can also nicely demonstrate
how so called pinch singularities \cite{GL99} are regulated within the
non-perturbative context of the thermalization process. 
These singularities do (and have to) appear in the
perturbative evaluation of higher order diagrams within the CTP description
of non-equilibrium quantum field theory. They are simply connected to
the standard divergence in elementary scattering theory.
The occurrence of pinch singularities signals the occurrence of (onshell) damping or
dissipation. This necessitates in the description of the evolution of
the system by means of non-perturbative transport equations.

As a further application
(discussed in the talk) for the direct use of semi-classical Langevin
equations we want to mention the recent work in \cite{PRL}:
By applying a microscopically motivated Langevin description
of the linear sigma model, one can investigate
the stochastic evolution of a so called disoriented chiral
condensate (DCC) in a rapidly expanding system, expected to occur
in ultrarelativistic heavy ion collisions.
Within such an approach
one finds that an experimentally feasible DCC, if it does exist
in nature, has to be a rare event,
but still occuring with some finite and nonvanishing probability.
The statistical distribution of final emitted pion number
out of domains shows a striking nonpoissonian and nontrivial
behaviour.
One should indeed interpret those
particular rarely occuring events as semi-classical `pion bursts' similar to
the mystique Centauro candidates.
A further analysis of this unusual
distribution by means of the cumulant expansion shows that the reduced
higher order factorial cumulants exhibit
an abnormal, exponentially increasing tendency
and thus serves as a new and powerful signature.
The occurence of a rapid chiral phase transition
(and thus DCCs) might then probably only be identified
experimentally by inspecting
higher order facorial cumulants $\theta _m$ ($m\ge 3$)
for taken distributions of low momentum pions.


\end{document}